\documentclass[12pt,preprint]{aastex}

\usepackage{emulateapj5,pstricks}

\newcommand{\be}{\begin{equation}}
\newcommand{\ee}{\end{equation}}
\newcommand{\ba}{\begin{eqnarray}}
\newcommand{\ea}{\end{eqnarray}}

\shorttitle{}
\shortauthors{Wang \& Mukherjee}

\begin{document}

\title{Model-Independent Constraints on Dark Energy Density\\
from Flux-averaging Analysis of Type Ia Supernova Data}
\author{Yun~Wang$^{1}$, and Pia~Mukherjee$^{1}$}
\altaffiltext{1}{Department of Physics \& Astronomy, Univ. of Oklahoma,
                 440 W Brooks St., Norman, OK 73019;
                 email: wang,pia@nhn.ou.edu}

\begin{abstract}

We reconstruct the dark energy density $\rho_X(z)$
as a free function from current type Ia supernova 
(SN Ia) data (Tonry et al. 2003; Barris et al. 2003; Knop et al. 2003), 
together with the Cosmic Microwave Background (CMB) shift parameter 
from CMB data (WMAP, CBI, and ACBAR), and 
the large scale structure (LSS) growth factor 
from 2dF galaxy survey data.
We parametrize $\rho_X(z)$ as a continuous function,
given by interpolating its amplitudes at equally spaced
$z$ values in the redshift range covered by SN Ia data,
and a constant at larger $z$ (where $\rho_X(z)$ is only weakly 
constrained by CMB data).
We assume a flat universe, and use the Markov Chain Monte Carlo 
(MCMC) technique in our analysis.
We find that the dark energy density 
$\rho_X(z)$ is constant for $0 \la z \la 0.5$ and
increases with redshift $z$ for $0.5 \la z \la 1$ at 
68.3\% confidence level, but is consistent with a constant at 95\% 
confidence level.
For comparison, we also give constraints on a constant
equation of state for the dark energy.

Flux-averaging of SN Ia data is required 
to yield cosmological parameter constraints that are free
of the bias induced by weak gravitational lensing \citep{Wang00b}.
We set up a consistent framework for flux-averaging
analysis of SN Ia data, based on \cite{Wang00b}.
We find that flux-averaging of SN Ia data leads
to slightly lower $\Omega_m$ and smaller time-variation 
in $\rho_X(z)$. This suggests
that a significant increase in the number of SNe Ia 
from deep SN surveys on a dedicated telescope \citep{Wang00a}
is needed to place a robust constraint on the
time-dependence of the dark energy density.

\end{abstract}



\section{Introduction}

Observational data of type Ia supernovae (SNe Ia) indicate that our universe 
is dominated by dark energy today
\citep{Riess98,Perl99}. 
The nature of dark energy is one of the great mysteries in cosmology at present.
The time-dependence of the dark energy density $\rho_X(z)$
can illuminate the nature of dark energy,
and help differentiate among the various dark energy
models (for example, \cite{fafm,peebles88,frieman,caldwell98,Dode00,
ddg,Albrecht02,bo02,freeselewis,Griest02,Sahni02,Carroll03,farrar03}.
See \cite{Pad03} and \cite{peebles03} for reviews with more complete 
lists of references).

SNe Ia can be calibrated to be good cosmological standard candles,
with small dispersions in their peak luminosity \citep{Phillips93,Riess95}.
The measurements of the distance-redshift relations of SNe Ia
are most promising for constraining the time variation of
the dark energy density $\rho_X(z)$.
The luminosity distance $d_L(z) = (1+z) r(z)$, with
the comoving distance $r(z)$ given by
\be
r(z)= c H_0^{-1} \int_0^z \frac{dz'}{E(z')},
\ee 
where
\be
\label{eq:E(z)}
E(z) \equiv \left[ \Omega_m(1+z)^3 + 
\Omega_k (1+z)^2 + \Omega_X \rho_X(z)/\rho_X(0) \right]^{1/2},
\ee
with $\Omega_k \equiv 1-\Omega_m-\Omega_X$.
If the dark energy equation of state $w_X(z)=w_0+w_1 z$, then
\be
\frac{\rho_X(z)}{\rho_X(0)}= e^{3 w_1 z}\, (1+z)^{3(1+w_0-w_1)}.
\ee
The dark energy density $\rho_X(z)=\rho_X(0)(1+z)^{3(1+w_0)}$
if the dark energy equation of state is a constant given by $w_0$.

Most researchers have chosen to study dark energy by
constraining the dark energy equation of state $w_X$.
However, due to the smearing effect \citep{MBS00}
arising from the multiple integrals 
relating $w_X(z)$ to the luminosity distance
of SNe Ia, $d_L(z)$,
it is extremely hard to constrain $w_X$ using SN data 
without making specific assumptions about $w_X$
\citep{Barger,Huterer01,Maor02,Wass02}.
If we constrain the dark energy density $\rho_X(z)$ instead,
we minimize the smearing effect by removing one 
integral \citep{Wang01a,Tegmark02,Daly03}.

It is important that 
there are a number of other probes of dark energy that are 
complementary to SN Ia data (for example, see \cite{Poda01,Schulz01,Bean02,Hu02,
Sereno02,Bern03,Huterer03,Jimenez03,Maju03,Mukherjee03,Munshi03,
Munshi03b,Seo03,Viel03,Weller03,Zhu03}).
Since different methods differ in systematic uncertainties,
the comparison of them will allow for consistency checks,
while the combination of them could yield tighter constraints
on dark energy (for example, see \cite{Gerke02,Hanne02,Kujat02}).

The most pressing question about dark energy that can be addressed
by observational data is whether the dark energy density varies
with time.
In order to constrain the time-variation of dark energy
density in a robust manner, it is important that we allow
the dark energy density to be an arbitrary function of redshift $z$
\citep{Wang01a,Wang01b,Wang03}. 
In this paper, we present a model-independent reconstruction
of the dark energy density $\rho_X(z)$, using
SN Ia data published recently by 
the High-$z$ Supernova Search Team (HZT) 
and the Supernova Cosmology Project (SCP)
\citep{Tonry03,Barris03,Knop03}, together with
constraints from
CMB (WMAP \citep{Bennett03}, CBI\citep{CBI}, and ACBAR\citep{ACBAR}) 
and large scale structure (LSS) data from
the 2dF galaxy survey \citep{Percival02}.

Note that for clarity of presentation, we will label the samples 
of SNe Ia that we use
according to the papers in which they were published.
Hence we will refer to the 194 SNe Ia from \cite{Tonry03}
and \cite{Barris03} as the
``Tonry/Barris sample'', and the 58 SNe Ia from
\cite{Knop03} as the ``Knop sample''.

Flux-averaging of SN Ia data is required 
to yield cosmological parameter constraints that are free
of the bias induced by weak gravitational lensing \citep{Wang00b}.
In this paper, we set up a consistent framework for flux-averaging
analysis of SN Ia data, based on \cite{Wang00b}.

We assume a flat universe, 
and use the Markov Chain Monte Carlo (MCMC) technique
in our analysis.

Sec.2 contains a consistent framework for flux-averaging analysis.
We present our constraints on dark energy in Sec.3.
Sec.4 contains a summary and discussions.

\section{A consistent framework for flux-averaging analysis}

Since our universe is inhomogeneous in matter distribution,
weak gravitational lensing by galaxies is one of the
main systematics\footnote{The others systematics are possible 
gray dust \citep{Aguirre99} and SN Ia peak luminosity evolution 
\citep{Drell00,Riess99,Wang00b};
so far, there is no clear evidence of either.} 
in the use of SNe Ia as cosmological standard candles 
\citep{Kantowski95,Frieman97,Wamb97,HolzWald98,ms99,Wang99,Wang02,Munshi03}.
Flux-averaging {\it justifies} the use
of the distance-redshift relation for a smooth universe
in the analysis of type Ia supernova (SN Ia) data \citep{Wang00b}.
Flux-averaging of SN Ia data is required 
to yield cosmological parameter constraints that are free
of the bias induced by weak gravitational lensing 
\citep{Wang00b}.\footnote{To avoid missing the faint end (which is
fortunately steep) of the magnification distribution of observed SNe Ia, only 
SNe Ia detected well above the threshold should be used in flux-averaging.}
Here we set up a consistent framework for flux-averaging
analysis of SN Ia data, based on \cite{Wang00b}.

\subsection{Why flux-averaging?}

The reason that flux-averaging can remove/reduce gravitational lensing bias
is that due to flux conservation, the average magnification of
a sufficient number of standard candles at the same redshift
is one. 

The observed flux from a SN Ia can be written as
\be
F(z)=  F_{int} \, \mu, \hskip 1cm 
F_{int}= F^{tr}(z|\mbox{\bf s}^{tr})+ \Delta F_{int},
\ee
where $F^{tr}(z|\mbox{\bf s}^{tr})$ is the predicted flux due to the 
true cosmological model parametrized by the set of cosmological 
parameters $\{ {\bf s}^{tr} \}$,
$\Delta F_{int}$ is the uncertainty in SN Ia peak brightness
due to intrinsic variations in SN Ia peak luminosity and
observational uncertainties, 
and $\mu$ is the magnification due to
gravitational lensing by intervening matter.
Therefore
\be
\Delta F^2 = \mu^2 \, \left(\Delta F_{int}\right)^2 + 
\left(F_{int}\right)^2 \left(\Delta \mu\right)^2.
\ee

Without flux-averaging, we have
\ba
\label{eq:chi2flux_true}
\chi^2_{N_{data}}(\mbox{\bf s}^{tr}) & = &\sum_i \frac{ \left[F(z_i) -
F^{tr}(z_i|\mbox{\bf s}^{tr})\right]^2}{\sigma_{F,i}^2} \nonumber \\
&=& \sum_i \frac{ \left[F^{tr}(z_i) \,(\mu_i-1)\right]^2+ 
\mu_i^2\left[ \Delta F_{int}^{(i)}\right]^2 
}{\sigma_{F,i}^2}+ \nonumber \\
& & \hskip 0.2cm 
2 \sum_i \frac{ F^{tr}(z_i) \Delta F_{int}^{(i)}\,\mu_i(\mu_i-1) 
}{\sigma_{F,i}^2} \nonumber \\
&=& N_{data} + 2 \sum_i \frac{ F^{tr}(z_i) \Delta F_{int}^{(i)}\,\mu_i(\mu_i-1) 
}{\sigma_{F,i}^2}.
\ea

The flux-averaging described in the Section 2.3
leads to the flux in each redshift bin
\be
\overline{F}(\overline{z}_{i_{bin}})= F^{tr}(\overline{z}_{i_{bin}})
\, \langle \mu \rangle_{i_{bin}} + \langle \mu \Delta F_{int}\rangle_{i_{bin}}.
\ee
For a sufficiently large number of SNe Ia in the $i$-th bin, 
$\langle \mu \rangle_{i_{bin}}$=1. Hence
\be
\label{eq:chi2bin}
\chi^2_{N_{bin}}(\mbox{\bf s}^{tr}) \simeq \sum_{i_{bin}}^{N_{bin}}
\frac{ \left[ \langle \mu \Delta F_{int}\rangle_{i_{bin}}\right]^2}
{\sigma^2_{F,i_{bin}}} 
\simeq \sum_{i_{bin}}^{N_{bin}}
\frac{ \left[ \langle\Delta F_{int}\rangle_{i_{bin}}\right]^2}
{\sigma^2_{F,i_{bin}}} 
< N_{bin}.
\ee
Comparison of Eq.(\ref{eq:chi2bin}) and Eq.(\ref{eq:chi2flux_true})
shows that flux-averaging can remove/reduce the gravitational
lensing effect, and
leads to a smaller $\chi^2$ per degree
of freedom for the true model,
compared to that from without flux-averaging.

\subsection{Flux statistics versus magnitude statistics}

Normally distributed measurement errors are required
if the $\chi^2$ parameter estimate is to be a
maximum likelihood estimator \citep{Press94}.
Hence, it is important that we use the $\chi^2$
statistics with an observable that has a error
distribution closest to Gaussian.

So far, it has been assumed that the distribution
of observed SN Ia peak brightness is Gaussian
in {\it magnitudes}. Therefore, for a given set
of cosmological parameters $\{ {\bf s} \}$ 
\be
\label{eq:chi2_mu0}
\chi^2 = \sum_i \frac{ \left[\mu_0(z_i) -\mu^p_0(z_i|{\bf s}) 
\right]^2}{\sigma_{\mu_0}^2},
\ee
where $\mu^p_0(z)=5\,\log\left(d_L(z)/\mbox{Mpc}\right)+25$,
and $d_L(z) = (1+z) r(z)$ is the luminosity distance.

However, while we do not have a very clear understanding
of how the intrinsic dispersions in SN Ia peak luminosity
is distributed, the distribution of observational uncertainties in
SN Ia peak brightness is Gaussian in {\it flux}, since
CCD's have replaced photometric plates as detectors of photons.

In this paper, we assume that the intrinsic dispersions
in SN Ia peak brightness is Gaussian in {\it flux}, and not
in magnitude as assumed in all previous publications.
The justifications for this preference will be presented in detail
elsewhere \citep{Wang04}\footnote{Our study of intrinsic
peak luminosities of nearby SNe Ia shows that their distribution is much more
Gaussian in flux than in magnitude.}. 
Thus,
\be
\label{eq:chi2flux}
\chi^2_{N_{data}}(\mbox{\bf s})  = \sum_i \frac{ \left[F(z_i) -
F^p(z_i|\mbox{\bf s})\right]^2}{\sigma_{F,i}^2}.
\ee
Since the peak brightness of SNe Ia have been given in magnitudes 
with symmetric error bars, $m_{peak}\pm \sigma_m$, we obtain 
equivalent errors in flux as follows:
\be
\sigma_F \equiv \frac{F(m_{peak}+\sigma_m)-F(m_{peak}-\sigma_m)}{2}
\ee

We will refer to Eq.(\ref{eq:chi2_mu0}) as ``magnitude statistics'',
and Eq.(\ref{eq:chi2flux}) as ``flux statistics''.
For reference and comparison, we present results in
both ``magnitude statistics'' and ``flux statistics''
in this paper. However, a consistent framework for
flux-averaging is only straightforward in ``flux statistics''.
\footnote{If the the dispersions in SN Ia peak brightness were Gaussian  
in magnitude, flux-averaging would introduce a small bias.}

\subsection{A recipe for flux-averaging}

The procedure for flux-averaging in \cite{Wang00b} is
for minimizing $\chi^2$ using the subroutines from
Numerical Recipes \citep{Press94}. 
As described in \cite{Wang00b}, the fluxes of SNe Ia in a redshift bin
should only be averaged {\it after} removing their redshift dependence,
which is a model-dependent process.
For $\chi^2$ statistics
using MCMC or a grid of parameters, 
here are the steps in flux-averaging:

(1) Convert the distance modulus of SNe Ia into 
``fluxes'',
\be
F(z_j) \equiv 10^{-(\mu_0(z_j)-25)/2.5} =  
\left( \frac{d_L^{data}(z)} {\mbox{Mpc}} \right)^{-2}.
\ee

(2) For a given set of cosmological parameters $\{ {\bf s} \}$,
obtain ``absolute luminosities'', \{${\cal L}(z_j)$\}, by
removing the redshift dependence of the ``fluxes'', i.e.,
\be
{\cal L}(z_j) \equiv d_L^2(z_j |{\bf s})\,F(z_j).
\ee

(3) Flux-average the ``absolute luminosities'' \{${\cal L}^i_j$\} 
in each redshift bin $i$ to obtain $\left\{\overline{\cal L}^i\right\}$:
\be 
 \overline{\cal L}^i = \frac{1}{N}
 \sum_{j=1}^{N} {\cal L}^i_j(z^i_j),
 \hskip 1cm
 \overline{z_i} = \frac{1}{N}
 \sum_{j=1}^{N} z^i_j. 
\ee

(4) Place $\overline{\cal L}^i$ at the mean redshift $\overline{z}_i$ of
the $i$-th redshift bin, now the binned flux is
\be
\overline{F}(\overline{z}_i) = \overline{\cal L}^i /
d_L^2(\overline{z}_i|\mbox{\bf s}).
\ee
The 1-$\sigma$ error on each binned data point $\overline{F}^i$,
$\sigma^F_i$, is taken to be the root mean square of the 1-$\sigma$ errors on the 
unbinned data points in the $i$-th redshift bin, \{$F_j^i$\} ($j=1,2,...,N$), 
multiplied by $1/\sqrt{N}$ (see \cite{Wang00a}).

(5) For the flux-averaged data, 
$\left\{\overline{F}(\overline{z}_i)\right\}$, we find
\be
\label{eq:chi2}
\chi^2 = \sum_i \frac{ \left[\overline{F}(\overline{z}_i) -
F^p(\overline{z}_i|\mbox{\bf s}) \right]^2}{\sigma_{F,i}^2},
\ee
where $F^p(\overline{z}_i|\mbox{\bf s}_1)=
\left( d_L(z|\mbox{\bf s}) /\mbox{Mpc} \right)^{-2}$.

\section{Constraints
on dark energy}

The Tonry/Barris SN Ia sample consists of
194 SNe Ia with $z>0.01$ and extinction $A_V<0.5$
\citep{Tonry03,Barris03}.
To examine the effect of the two SNe Ia at the high
reshift end ($z=1.199$ and $z=1.755$ respectively),
we also present the results for
193 SNe Ia (omitting the SN Ia at $z=1.755$)
and 192 SNe Ia (omitting the two SNe Ia at $z=1.199$ and $z=1.755$).

The Knop SN Ia sample consists of 58 SNe Ia
(the ``All SCP SNe'' data set from \cite{Knop03}).
These data should be compared with
$m_B^{eff}=5 \log\left(H_0 d_L\right) + \mbox{offset}$, with
\be
\label{eq:offset}
\mbox{offset}=5 \,\log\left(2997.9/h\right)+25 + M_{SN},
\ee
where $M_{SN}$ is the peak absolute magnitude of SNe Ia.
Note that for the Knop sample, the flux statistics must
be done with a revised definition of flux,
\be
F(z)=\left( \frac{\left[H_0\,d_L(z)\right]^{data}}{{\mbox{Mpc}}}\right)^{-2}
= 10^{-2 \left(m_{eff}^B-\mbox{offset}\right)/5},
\ee
to be compared with theoretical prediction
of $F^p(z|\mbox{\bf s})=\left[ H_0\,d_L(z|\mbox{\bf s})/
{\mbox{Mpc}} \right]^{-2}$ for a given set of cosmological
parameters $\{\mbox{\bf s}\}$.

Note that an additional uncertainty from the
redshift dispersion due to peculiar velocity
must be added to the uncertainty of
each SN Ia data point. The Knop sample
already include a dispersion of 300~km/s along the
line of site. To add 500~km/s dispersion in $z$ to
the SN data in the Tonry/Barris sample,
one must propagate $\sigma_z= c^{-1} 500$~km/s into an
additional uncertainty in the luminosity distance $d_L(z)$,
then add it in quadrature to published uncertainty in $d_L(z)$.
Note that this process is {\it dependent} on the cosmological
model, and must be done for each set of cosmological parameters
during the likelihood analysis \citep{Riess98,Wang00b}.

To obtain tighter constraints on dark energy,
we also include constraints from CMB (WMAP, CBI, ACBAR)
and LSS (2dF) in our analysis. Since CMB data clearly
indicate that we live in a flat universe, we
assume $\Omega_m+\Omega_X=1$
in all our results.\footnote{Allowing both $\Omega_m$ and $\Omega_X$
to vary would lead to greatly increased uncertainty
in dark energy constraints, such that no interesting
constraints can be obtained from current data.}

We use the Markov Chain Monte Carlo (MCMC) technique (see
\cite{neil} for a review),
illustrated for example in \cite{LB02}, in the likelihood analysis.
At its best, the MCMC method scales approximately linearly in computation
time with the number of parameters. 
The method samples from the full posterior distribution of the
parameters, and from these samples the marginalized posterior distributions
of the parameters can be estimated. We have derived all our 
probability distribution functions (pdf) of the cosmological parameters from
$10^6$ MCMC samples.
 
\subsection{The likelihood analysis}

We use a $\chi^2$ statistic
\be
\chi^2 = \chi^2_{SN}+ \chi^2_{CMB}+ \chi^2_{LSS},
\ee
where $\chi^2_{SN}$ is given by Eq.(\ref{eq:chi2}) and (\ref{eq:chi2flux})
for flux statistics (with and without flux-averaging), and Eq.(\ref{eq:chi2_mu0}) 
for magnitude statistics. $\chi^2_{CMB}$ and $\chi^2_{LSS}$ are contributions
from CMB and LSS data respectively.
The likelihood ${\cal L} \propto e^{-\chi^2/2}$
if the measurement errors are Gaussian \citep{Press94}.

When the cosmological parameters are varied, 
the shift in the whole CMB angular spectrum is
determined by the shift parameter \citep{Bond97,Mel02,Odman02}
\be
\label{eq:R}
{\cal R} = \sqrt{\Omega_m}\, H_0 \, r(z_{dec})
\ee
where $r(z_{dec})$ denotes the comoving distance to the decoupling
surface in a flat universe.
Note that this is a robust way to include CMB constraints
since the CMB depends on $\Omega_m$ and $h$ in the combination of the
physical parameter $\Omega_m h^2$.
The results from CMB (WMAP, CBI, ACBAR)
data correspond to ${\cal R}_0 = 1.716 \pm 0.062$
(using results in \cite{Spergel03}).
We include the CMB data in our analysis by adding 
$\chi^2_{CMB}= \left[({\cal R}-{\cal R}_0)/\sigma_{\cal R}\right]^2$,
where ${\cal R}$ is computed for each model using
Eq.(\ref{eq:R}).

Following \cite{Knop03}, we include the LSS constraints from
2dF in terms of the growth parameter $f=d\mbox{ln}D/d\mbox{ln}a$,
where $a$ is the cosmic scale factor, and
$D$ is the linear fluctuation growth factor,
$D(t)=\delta^{(1)}(\mbox{\bf x}, t) /\delta(\mbox{\bf x})$, 
given by
\be
\label{eq:D}
\ddot D(t) + 2H(z) \dot D(t) - \frac{3}2{}\, \Omega_m H_0^2  (1+z)^3 D(t) =0,
\ee
where the dots denote derivatives with respect to $t$.
 
The Hubble parameter $H(z)=H_0\,E(z)$ (see Eq.(\ref{eq:E(z)})). 
Since $\beta= f/b_1$, the 2dF constraints of $\beta(z \sim 0.15)=0.49\pm 0.09$
\citep{Hawkins02} and $b_1=1.04 \pm 0.11$ \citep{Verde01}
yields $f_0 \equiv f(z=0.15) =0.51\pm 0.11$.
We include the 2dF constraints in our analysis by adding 
$\chi^2_{LSS}= \left\{ [f(z=0.15)-f_0]/\sigma_{f_0}\right\}^2$,
where $f=d\mbox{ln}D/d\mbox{ln}a$ is computed for each model using
$D$ obtained by numerically integrating Eq.(\ref{eq:D}).

Note that we have chosen to use only the most conservative and
robust information, the CMB shift parameter and the LSS growth factor, 
from CMB and LSS observations\footnote{These
observations provide a vast amount of information as
detailed in the publications from the WMAP and 2dF teams.}.
It is important that the limits on these are independent of the
assumption on dark energy made in the CMB and LSS data analysis.
Further, by limiting the amount of information that we
use from CMB and LSS observations to complement the SN Ia data,
we minimize the effect of the systematics inherent in
the CMB and LSS data on our results.

\subsection{Constraints
on a constant dark energy
equation of state $w_0$}

The most popular and simplest assumption about dark energy
is that it has a constant equation of state $w_0$.
Here we present constraints on a constant dark energy
equation of state. 

Fig.1 shows the marginalized pdf of the matter density fraction $\Omega_m$,
the dimensionless Hubble constant $h$, and the constant dark energy
equation of state $w_0$.

The first four rows of figures in Fig.1 are results obtained using
the Tonry/Barris SN Ia sample,
requiring that $z>0.01$ and extinction $A_V<0.5$
(which yields a total of 194 SNe Ia).
To examine the effect of two SNe Ia at the high
reshift end ($z=1.199$ and $z=1.755$ respectively),
we also present the results for
193 SNe Ia (omitting the SN Ia at $z=1.755$)
and 192 SNe Ia (omitting the two SNe Ia at $z=1.199$ and $z=1.755$).
The solid, dotted, and dashed lines indicate the results
for 192, 193, and 194 SNe Ia respectively. 

The first two rows of figures in Fig.1 are results for SN Ia data
from the Tonry/Barris sample 
only, without (first row) and with (second row) flux-averaging
($\Delta z=0.05$).
Note that inclusion of the two highest redshift SNe Ia
at $z=1.199$ and $z=1.755$ leads slightly higher $\Omega_m$
and more negative $w_0$.
Flux-averaging leads to broader pdf for $\Omega_m$,
with somewhat lower mean $\Omega_m$,
and broader pdf for $h$.

The 3rd and 4th rows of figures in Fig.1 are results for SN Ia data
from the Tonry/Barris sample, combined with constraints from CMB (WMAP, CBI, ACBAR)
and LSS (2dF) data.
The inclusion of the two highest redshift SNe Ia
at $z=1.199$ and $z=1.755$ makes less difference
in the estimated parameters, since the inclusion
of the CMB and LSS data reduces the relative weight 
of these two data points. The main effect of
flux-averaging is a broader pdf for $h$.

The 5th row of figures in Fig.1 are results obtained using
58 SNe Ia from the Knop sample,
using flux-averaged statistics (solid)
and magnitude statistics (dotted) respectively. 
Note flux averaging significantly broadens all the pdf's.
The central figure is equivalent to
a pdf in $h$ (see Eq.(\ref{eq:offset}).
These are consistent with similar results derived
using the SNe Ia from the Tonry/Barris sample (3rd row of figures
in Fig.1).

Table 1 gives the marginalized 68.3\% and 95\% 
confidence level (C.L.) 
of $\Omega_m$, $h$, and $w_0$. These have been
computed using $10^6$ MCMC samples.

\subsection{Constraints
on dark energy density as a free function}

To place model-independent constraints on dark energy,
we parametrize $\rho_X(z)$ as a continuous function,
given by interpolating its amplitudes at equally spaced
$z$ values in the redshift range covered by SN Ia data
($0 \leq z \leq z_{max}$),
and a constant at larger $z$ ($z>z_{max}$,
where $\rho_X(z)$ is only weakly constrained by CMB data).
The values of the dimensionless dark energy density
$f_i \equiv \rho(z_i)/\rho_X(0)$ ($i=1,2,...,n_{f})$ are the independent variables
to be estimated from data.
We interpolate $\rho_X(z)$ using a polynomial of order
$n_{f}$ for $0 \leq z \leq z_{max}$.

Since the present data can not constrain $\rho_X(z)$ for $n_f>2$,
we present results for $n_f=2$, i.e., with $\rho_X(z)$
parametrized by its values at $z=z_{maz}/2$, $z_{maz}$.

Fig.2 shows the marginalized pdf of the matter density fraction $\Omega_m$
(column 1), the dimensionless Hubble constant $h$ (column 2), 
and dimensionless
dark energy density at $z=z_{max}/2$ and $z=z_{max}$ (columns 3 and 4),
obtained using current SN Ia data \citep{Tonry03,Barris03,Knop03}, 
flux-averaged and combined with CMB (WMAP, CBI, ACBAR) and 2dF data. 
The first three rows of figures are results for
192, 193, and 194 SNe Ia from the Tonry/Barris sample,
while the fourth row are results for 58 SNe Ia from the Knop sample. 
The dark energy density at $z=z_{max}$ is not well constrained
when the $z=1.755$ SN Ia is included in the analysis;
this is as expected since this extends $\rho_X(z)$ to
$z_{max}=1.755$, with only one SN Ia at $z>1.2$. 

Note that when the estimated parameters are well constrained,
flux averaging generally leads to slightly lower estimates
of $\Omega_m$ and $\rho_X(z)$ (at $z=z_{max}/2$ and $z=z_{max}$).

Table 2 gives the marginalized 68.3\% and 95\% C.L. 
of $\Omega_m$, $h$, $\rho_X(z_{max}/2)/\rho_X(0)$,
and $\rho_X(z_{max})/\rho_X(0)$. These have been
computed using $10^6$ MCMC samples. 

Fig.3 shows the dark energy density $\rho_X(z)$ reconstructed 
from current SN Ia (the Tonry/Barris sample and the Knop sample), 
CMB (WMAP, CBI, ACBAR) and LSS (2dF) data. 
The heavy (light) lines indicate the 68.3\% (95\%) C.L. of the
reconstructed $\rho_X(z)$.
The dot-dashed line indicates the cosmological constant
model, $\rho_X(z)/\rho_X(0)=1$.
The 68.3\% and 95\% C.L.'s of $\rho_X(z)$ are marginalized confidence levels,
computed at each $z$ using $10^6$ MCMC samples, with the correlation
between $\rho_X(.5\,z_{max})$ and $\rho_X(z_{max})$ fully included.

Fig.3(a) shows the reconstructed $\rho_X(z)$
using 192, 193, and 194 SNe Ia from the Tonry/Barris sample, 
flux-averaged and combined with CMB (WMAP, CBI, ACBAR) and LSS (2dF) data. 
The densely (sparsely) shaded regions are the
68.3\% (95\%) C.L. of $\rho_X(z)$ for 192 SNe Ia
(at $z\le 1.056$).
The heavy (light) dotted and dashed lines
are the 68.3\% (95\%) C.L. of $\rho_X(z)$ for
193 and 194 SNe Ia respectively.
Note that the $\rho_X(z)$ reconstructed from 193 SNe Ia
(adding the SN Ia at $z=1.199$) nearly overlaps
from that from 192 SNe Ia for $z\la  1.056$.
However, the $\rho_X(z)$ reconstructed from 194 SNe Ia
(adding the SNe Ia at $z=1.199$ and $z=1.755$) deviates
notably from that from 192 SNe Ia for $ 0.7 \la z\la  1.056$,
although the 68.3\% C.L. regions overlap.
Clearly, the reconstructed $\rho_X(z)$ is constant for $0 \la z \la 0.5$ and
increases with redshift $z$ for $0.5 \la z \la 1$ at 
68.3\% C.L., but is consistent with a constant at 
95\% C.L. We note that at 90\% C.L.,
$\rho_X(.5\,z_{max})/\rho(0)=[0.83, 1.59]$,
and $\rho_X(z_{max})/\rho(0)=[1.03, 6.85]$;
this indicates that $\rho_X(z)$ varies with time
at approximately 90\% C.L.

Fig.3(b) shows the reconstructed $\rho_X(z)$
using 192 SNe Ia from the Tonry/Barris sample (same as in Fig.3(a)) and that
from the 58 SNe Ia of the Knop sample (dotted lines).
The 68.3\% C.L. regions overlap. However,
the 58 SCP SNe Ia seem to favor 
$\rho_X(z) \la 1$ at $0.5 \la z \la 1$,
and has much larger uncertainties at $ z \ga 0.5$.

\subsection{Comparison with previous work}

The SN observational teams have published their data
together with constraints on a constant dark energy
equation of state \citep{Tonry03,Knop03}. 
These results should be compared with our results for a 
constant $w_X(z)$ using magnitude statistics (see Table 1).
Using a 2dF prior of $\Omega_m h=0.20 \pm 0.03$
\citep{Percival02} and assuming a flat universe,
\cite{Tonry03} found that $-1.48 < w_0 < -0.72$ 
at 95\% C.L.;
this is close to $-1.24 < w_0 < -0.74$ at 95\% C.L.
(using 192 SNe Ia together with CMB and LSS constraints
as discussed in Sec.3.1) from Table 1.
Using the same CMB and LSS constraints as us
and assuming a flat universe,
\cite{Knop03} found that $w_0= -1.05 ^{+0.15}_{-0.20}$
(statistical) $\pm 0.09$ (identified systematics)
at 68.3\% C.L.; close to our results of 
$w_0 = -0.99 \pm 0.16$ at 68.3\% C.L.

At the completion of our analysis, 
we became aware that \cite{Alam03} and \cite{Choud03} have found that
current SN Ia data favor $w_X(z)<-1$.
Although our results are qualitatively consistent with these,
there are significant differences in both analysis technique and
quantitative results.

\cite{Alam03} used 172 SNe Ia from \cite{Tonry03},
and obtained a reconstructed $w_X(z)$ that deviates 
quite significantly from $w_X(z)=-1$. In their paper (v2),
Figs.3 and 14 (both assuming $\Omega_m=0.3$) 
are consistent with what we found.
We note that some of their reconstructed $w_X(z)$'s (see
their Figs.4, 6, 8, 10, and 16) have decreasing 
errors for $z\ga 1.2$ (where there are only two SNe Ia).
This illustrates the fact that while free fitting forms for 
$\rho_X(z)$ or $w_X(z)$ generally give increasing errors with 
redshift (large error where there are very few observed SNe Ia, 
see Fig.3 of this paper), this may not be true for other 
parametrizations (such as used in \cite{Alam03}).

\cite{Choud03} used 194 SNe Ia (the Tonry/Barris sample),
and presented their results for $w_X(z)=w_0 + w_1\,z/(1+z)$
with $\Omega_m=0.29$, 0.34, and 0.39 (no marginalization
over $\Omega_m$).

\section{Summary and Discussion}

In order to place model-independent constraints on dark energy,
we have reconstructed the dark energy density $\rho_X(z)$
as a free function from current SN Ia data (Tonry et al. 2003; 
Barris et al. 2003; Knop et al. 2003), 
together with CMB (WMAP, CBI, and ACBAR) and LSS (2dF) data.
We find that the dark energy density 
$\rho_X(z)$  is constant for $0 \la z \la 0.5$ and
increases with redshift $z$ for $0.5 \la z \la 1$ at 
68.3\% C.L., but is consistent with a constant at 95\% C.L.
(see Fig.3). 

Flux-averaging of SN Ia data is required 
to yield cosmological parameter constraints that are free
of the bias induced by weak gravitational lensing \citep{Wang00b}.
We have developed a consistent framework for flux-averaging
analysis of SN Ia data, and applied it to current SN Ia data.
We find that flux-averaging of SN Ia data generally leads
to slightly lower $\Omega_m$ and smaller time-variation 
in $\rho_X(z)$. 

We note that flux-averaging of SNe Ia has more effect
on the Knop sample than the Tonry/Barris sample.
This may be due to the fact that the measurement errors of 
the majority of the SNe Ia in the Tonry/Barris sample have been 
``Gaussianized'' in magnitudes by averaging over several different analysis
techniques \citep{Tonry03}. However, it is likely that SN Ia peak brightness
distribution is Gaussian in {\it flux}, instead of 
magnitudes \citep{Wang04}.
A consistent framework for flux-averaging is only
straightforward (as presented in Sec.2) if the
distribution of SN Ia peak brightnesses
is Gaussian in flux.
Our results suggest that observers should publish observed SN Ia peak 
brightnesses with uncertainties in flux, to allow 
detailed flux-averaging studies.

Our results include an estimate of the Hubble constant
$H_0=h\, 100\,$km/s$\,$Mpc$^{-1}$ from the Tonry/Barris sample
of 194 SNe Ia. Since the Tonry/Barris sample data used
a fixed value of $h_{fix}=0.65$ \citep{Tonry03}
in the derived distances, we divide their derived
distances $H_0 \, d_L(z)$ by $h_{fix}$ and marginalize
over $H_0$ in our analysis. Our MCMC method yields
smooth pdf's for all marginalized parameters.  
The errors on the estimated $h$ in Tables 1 \& 2
are statistical errors only, not including a much larger
systematic error contributed by the intrinsic dispersion
in SN Ia peak luminosity of $\sigma_m^{int} \simeq 0.17$
magnitudes \cite{H96}. This implies a systematic uncertainty 
in $h$ of 7.83\%, or $\sigma_h^{int} \simeq 0.05$ for the
$h$ values tabulated in Table 1 \& 2.
This yields an estimate for $h$ that overlap with those
from \cite{Branch98} and \cite{Freedman01} within 1~$\sigma$.

It is intriguing that the current SN Ia data, together with
CMB and galaxy survey data, indicate that $\rho_X(z)$ varies with time
at approximately 90\% C.L. (see Fig.3 and Sec.3.3).
If the trend in $\rho_X(z)$ that we have found is confirmed by future
observational data, it will have revolutionary implications
for particle physics and cosmology.
Since the uncertainty in $\rho_X(z)$
is large where there are few observed SNe Ia (see Fig.3),
we expect that a significant increase in the number of SNe Ia,
obtained from dedicated deep SN Ia searches \citep{Wang00a},
will allow us to place robust and more stringent constraints on the
time-dependence of the dark energy density.

\acknowledgements
This work is
supported in part by NSF CAREER grant AST-0094335. 
We are grateful to David Branch and Michael Vogeley
for helpful comments.

\begin{center}
Table 1\\
{\footnotesize{Estimated cosmological parameters (mean, 68.3\% C.L., 95\% C.L.)
assuming $w_X(z)=w_0$}}

{\footnotesize
\begin{tabular}{lcccc}
\hline 
\hline
{\bf Tonry/Barris sample SNe} &  & & & \\
& $\Omega_m$ & $h^a$ & $w_0$ & $\chi^2_{min}$/$N_{dof}^c$ \\
\hline 
{\bf 192 SNe} ($z_{max}$=1.056) & & &&\\
flux, binned$^b$ & .47 [.36, .57][.14, .63]  & .661 [.645, .674][.631, .690] & 
 -2.37 [-3.56, -1.20][-5.34, -.69] & 13.34/19  \\
flux, unbinned  & .47 [.43, .52][.34, .55] & .656 [.644, .666][.636, .676] & 
-3.08 [-4.09, -2.08][-5.50, -1.46]  & 209.10/189 \\
mag.; unbinned  & .49 [.41, .57][.22, .61] & .662 [.651, .671][.644, .681] & 
-2.25 [-3.18, -1.34][-4.36, -.80] & 193.36/189 \\
\hline
{\bf 193 SNe} ($z_{max}$=1.199) & & &&\\
flux, binned$^b$    & .48 [.39, .58][.18, .63] & .661 [.645, .676][.631, .690] &
-2.53 [-3.85, -1.30][-5.54, -.73] &  14.36/20\\
flux, unbinned  & .48 [.43, .52][.35, .56] & .656 [.645, .665][.637, .676] & 
-3.13 [-4.10, -2.13][-5.64, -1.50]   & 210.56/190 \\
mag.; unbinned   & .51 [.44, .58][.29, .62]  & .663 [.653, .671][.644, .683] &  
-2.47 [-3.42, -1.49][-5.14, -.91] & 194.86/190 \\
\hline
{\bf 194 SNe} ($z_{max}$=1.755) & & &&\\
flux, binned$^b$    & .49 [.40,  .58][.20, .63]  & .661 [.645, .675][.632, .690] &  
 -2.54 [-3.80, -1.34][-5.62, -.76] & 14.40/21 \\  
flux, unbinned  & .48 [.44, .53][.37, .56]  & .656 [.645, .665][.637, .675] & 
 -3.17 [-4.13, -2.22][-5.47, -1.58] & 210.74/191 \\ 
mag.; unbinned  & .51 [.45, .58][.30, .62]  & .663 [.653, .671][.644, .683]  &  
-2.51 [-3.46, -1.55][-4.97, -.95] & 194.88/191 \\    
\hline 
\hline
{\bf 192 SNe} ($z_{max}$=1.056) {\bf + CMB \& LSS} &  & &&\\
flux, binned$^b$ & .28 [.23, .33][.19, .39]  & .652 [.638, .665][.627, .677] & 
-.95 [-1.09, -.82][-1.27, -.72]  &  15.44/21  \\
flux, unbinned  & .26 [.22, .31][.18, .36] & .643 [.634, .650][.627, .657] & 
 -1.15 [-1.29, -1.00][-1.53, -.90] &  216.76/191 \\
mag.; unbinned  & .29 [.24, .34][.21, .39] & .654 [.646, .661][.639, .668] & 
-.95 [-1.07, -0.83][-1.24, -.74] & 196.60/191 \\
\hline
{\bf 193 SNe} ($z_{max}$=1.199) {\bf + CMB \& LSS} &  & &&\\
flux, binned$^b$    & .29 [.24, .34][.20, .39]  & .652 [.638, .663][.626, .676] &
 -.95 [-1.08, -.82][-1.27, -.71] &  17.04/22 \\
flux, unbinned  & .27 [.22, 31][.18, .37] & .643 [.635, .650][.627, .658] & 
 -1.15 [-1.29, -1.00][-1.54, -.90] &  219.00/192 \\
mag.; unbinned   & .30 [.25, .35][.21, .40]  & .654 [.645, .660][.638, .667] &  
-.95 [-1.07, -.82][-1.25, -.74] & 199.04/192 \\
\hline
{\bf 194 SNe} ($z_{max}$=1.755) {\bf + CMB \& LSS} &  & &&\\
flux, binned$^b$    &  .29 [.24, .34][.20, .40] &  .651 [.638, .663][.625, .676] &  
 -.95 [-1.08, -.81][-1.28, -.71] &  17.44/23 \\  
flux, unbinned  & .27 [.22, .32][.18, .37]  & .642 [.634, .650][.627, .657] & 
-1.15 [-1.31, -1.00][-1.57, -.90] &  219.8/193 \\
mag.; unbinned   & .30 [.25, .35][.21, .41]  & .654 [.645, .660][.638, .668]  &  
 -.95 [-1.07, -.82][-1.27, -.73] & 199.50/193 \\     
\hline 
\hline
{\bf 58 Knop sample SNe} ($z_{max}$=0.863) &  & &&\\
\hskip 0.2cm  {\bf + CMB \& LSS} & $\Omega_m$ & offset-23.5 & $w_0$ & $\chi^2_{min}$/$N_{dof}^c$\\
\hline
flux, binned$^b$ & .22 [.15, .28][.11, .36]	& .414 [.347, .480][.286, .547]	&
-1.20 [-1.45, -.96][-1.73, -.74]	& 9.1/13\\
flux, unbinned  & .21 [.16, .26][.13, .32] & .395 [.356, .434][.321, .471] & 
 -1.18 [-1.34, -1.02][-1.54, -.88] &   61.34/57  \\
mag.; unbinned  & .27 [.21, .32][.17, .39] & .367 [.331, .406][.295, .441] & 
-.99 [-1.15, -.83][-1.34, -.70] & 55.86/57 \\
\hline
\end{tabular}
}
\end{center}
\footnotesize{$^a$ Statistical error only, not including the contribution
from the much larger SN Ia absolute magnitude error of $\sigma_h^{int}\simeq 0.05$
(see Sec.4).\\
$^b$ Flux-averaged with ${\Delta}z=.05$.\hskip 1cm
$^c$ The number of degrees of freedom.}

\begin{center}
Table 2\\
{\footnotesize{Estimated cosmological parameters (mean, 68.3\% C.L., 95\% C.L.)
for arbitrary $\rho_X(z)$}, from SN Ia data combined with CMB and LSS data}

{\footnotesize
\begin{tabular}{lccccc}
\hline 
\hline
& $\Omega_m$ & $h^a$ & $\rho_X(.5\,z_{max})/\rho_X(0)$ & $\rho_X(z_{max})/\rho_X(0)$ & 
$\chi^2_{min}$/$N_{dof}^c$\\
\hline 
& & {\bf Tonry/Barris sample} &&&\\
192 SNe &  ($z_{max}$=1.056) &  &  & & \\
binned flux$^b$  & .33 [.27,.39][.22, .46]  & .660 [.644, .673][.630, .688] & 
1.19 [.97, 1.42][.76, 1.67] & 3.61 [1.84, 5.41][.73, 7.53] & 13.28/20\\
unbinned flux  & .34 [.28, .39][.24, .45] & .655 [.645, .663][.637, .671] & 
 1.09 [.88, 1.31][.67, 1.56] & 5.02 [3.27, 6.82][1.98, 9.15]  & 208.26/190\\
 unbinned mag.   & .34 [.28, .40][.23, .46] & .662 [.652, .670][.645, .678] & 
1.22 [1.01, 1.43][.80, 1.66]&3.75 [2.05, 5.40][.94, 7.76]& 193.30/190 \\
\hline
193 SNe & ($z_{max}$=1.199) &  &  & & \\
binned flux$^b$ & .35 [.28, .41][.23, .48]  & .660 [.645, .674][.631, .688] &  
 1.39 [1.08, 1.69][.84, 2.10] & 4.95 [2.55, 7.34][.88, 10.35] & 14.24/21 \\
unbinned flux & .36 [.30, .42][.25, .49]  & .656 [.645, .664][.637, .672] & 
1.44 [1.06, 1.82][.80, 2.41] & 7.50 [4.39, 10.69][2.57, 15.57] &  209.42/191 \\
unbinned mag.  & .36 [.30, .42][.25, .48] & .662 [.653, .670][.645, .678]  & 
 1.42 [1.13, 1.71][.89, 2.08] & 5.14 [2.88, 7.43][1.41, 10.62] & 194.50/191 \\
\hline
194 SNe & ($z_{max}$=1.755) &  &  & & \\
binned flux$^b$  & .40 [.32, .48][.25, .54] & .661 [.646, .675][.632, .688] & 
 3.26 [1.76, 4.76] [1.02, 5.75] & 15.64 [6.22, 25.30] [.92, 30.53] &  14.5/22 \\
  unbinned flux  & .38 [.32, .43][.27, .48] & .654 [.644, .662][.637, .670] & 
 2.85 [1.88, 3.81][1.18, 4.69] & 14.78 [8.77, 20.58][4.36, 25.83] &  209.88/192 \\
 unbinned mag.  & .38 [.31, .44][.26, .50] & .662 [.652, .669][.645, .678] & 
 2.48 [1.66, 3.54][1.22, 4.35] & 10.60 [5.41, 17.81][2.68, 21.80] & 194.70/192  \\
\hline
\hline
& & {\bf Knop sample} &&&\\
58 SNe &  ($z_{max}$=0.863) &  &  & & \\
& $\Omega_m$ & offset-23.5 & $\rho_X(.5\,z_{max})/\rho(0)$ & 
$\rho_X(z_{max})/\rho(0)$ & $\chi^2_{min}$/$N_{dof}$\\
\hline
binned flux$^b$  & .34 [.25, .43][.18, .52]  & .311 [.210, .409][.127, .509] & 
.91 [.59, 1.22][.38, 1.61] & 5.92 [2.13, 9.73][.58, 13.21] &   7.2/12 \\
unbinned flux & .26 [.18, .33][.12, .41] & .381 [.331, .428][.286, .474] &
.85 [.67, 1.03][.50, 1.21] & 1.96 [.53, 3.55][.09, 5.45] & 61.32/56 \\
unbinned mag.  & .39 [.30 .48][.22, .56] & .309 [.256, .362][.208, .412] & 
1.18 [.89, 1.46][.67, 1.82]  & 6.10 [2.47, 9.59][.86, 12.57] & 53.90/56\\
\hline
\end{tabular}
}
\end{center}
\footnotesize{$^a$ Statistical error only, not including the contribution
from the much larger SN Ia absolute magnitude error of $\sigma_h^{int}\simeq 0.05$
(see Sec.4).\\
$^b$ flux-averaged with ${\Delta}z=.05$.\hskip 1cm
$^c$ The number of degrees of freedom.}

\setcounter{figure}{0}
\plotone{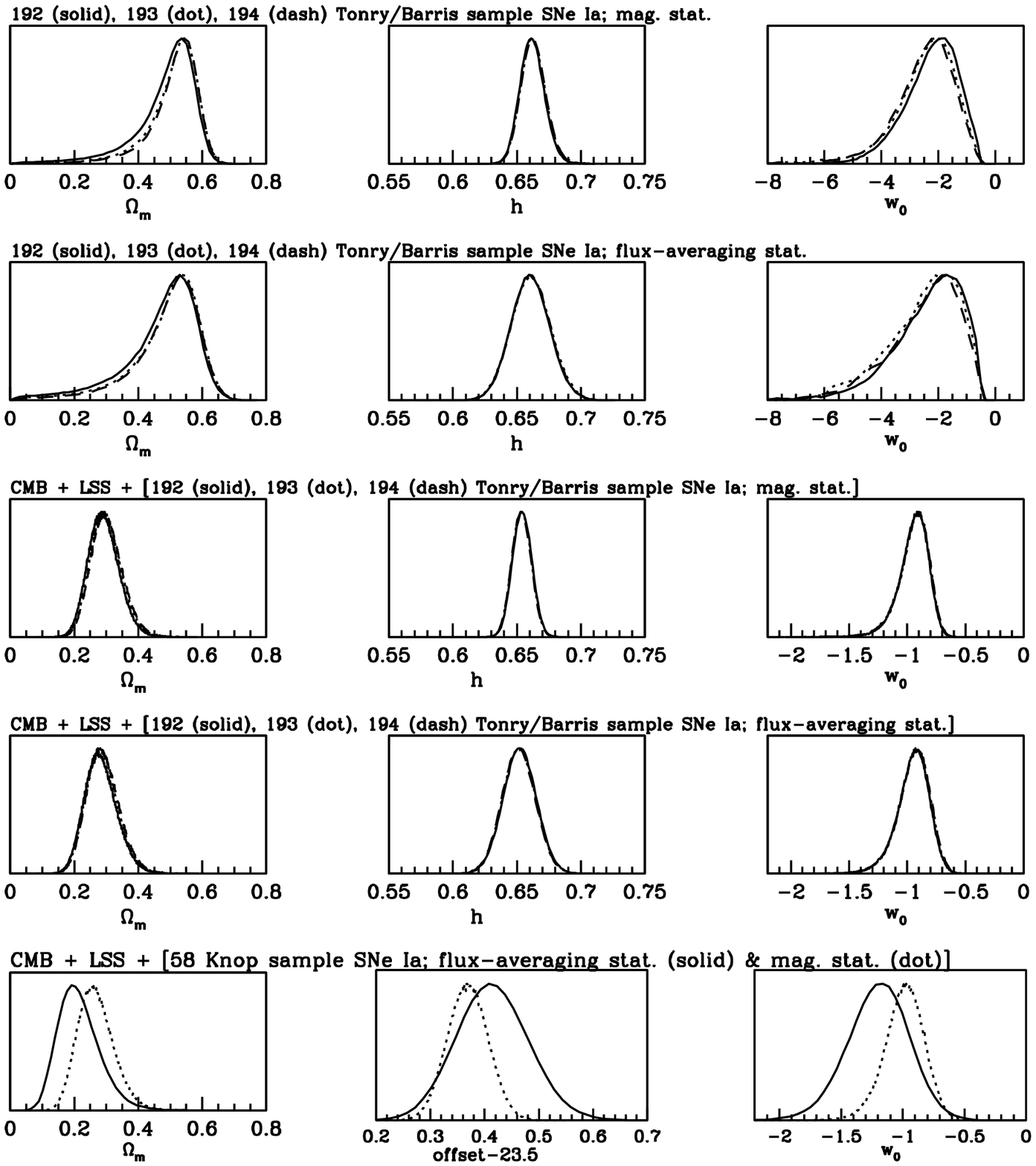}
\figcaption[f1.eps]
{The marginalized probability distributions of $\Omega_m$,
$h$, and the constant
equation of state for the dark energy $w_0$.}

\plotone{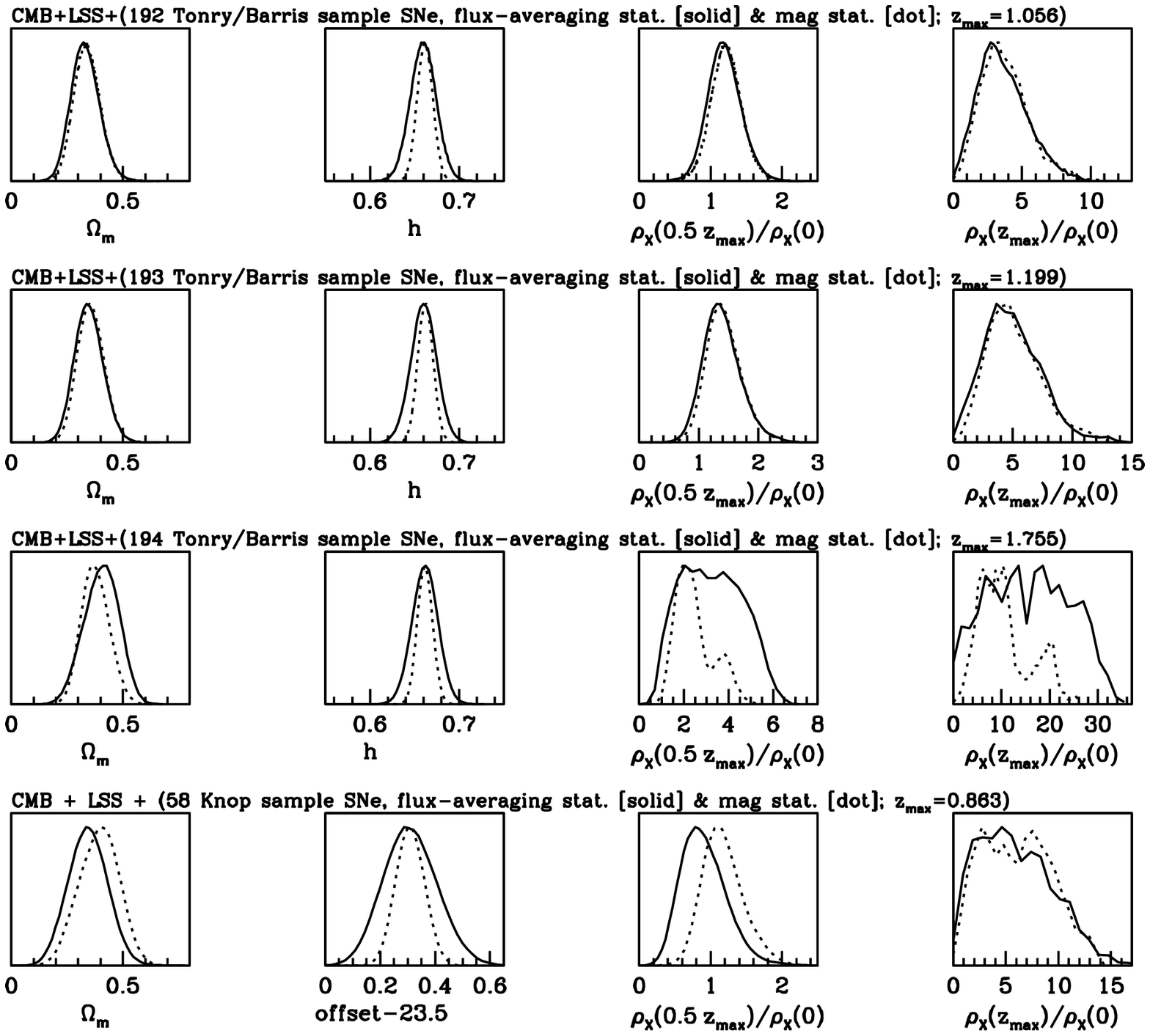}
\vskip -3cm 
\figcaption[f2.eps]
{The marginalized probability distributions of $\Omega_m$,
$h$, and dimensionless
dark energy density at $z=z_{max}/2$ and $z=z_{max}$.}

\plotone{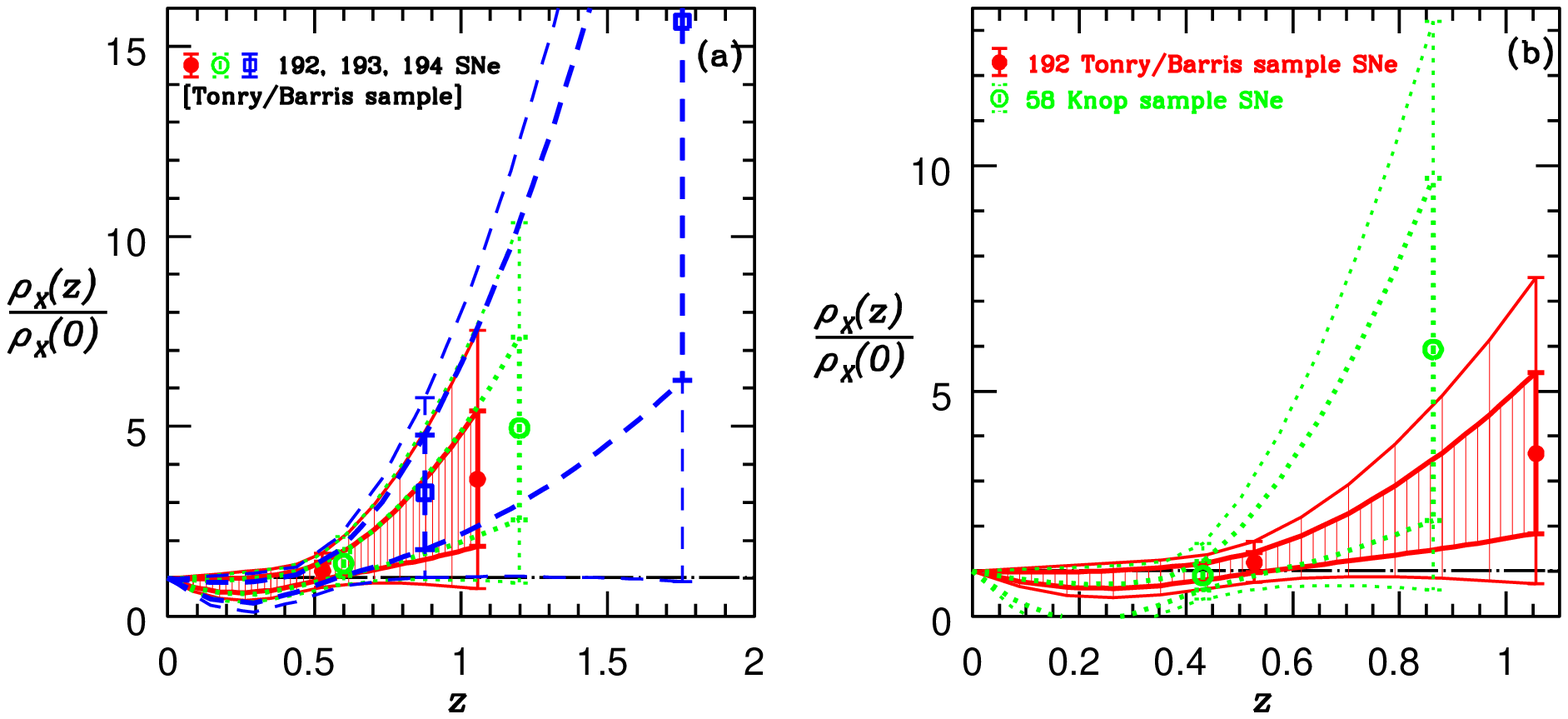}
\vskip -9cm 
\figcaption[f3.eps]
{The dark energy density $\rho_X(z)$ reconstructed from current SN Ia (Tonry/Barris
sample and Knop sample), CMB (WMAP, CBI, ACBAR), and LSS (2dF) data. 
The densely (sparsely) shaded regions are the
68.3\% (95\%) C.L. of $\rho_X(z)$ for 192 Tonry/Barris sample SNe Ia
(at $z\le 1.056$).
The heavy (light) lines indicate the 68.3\% (95\%) C.L. of the
reconstructed $\rho_X(z)$.
The dot-dashed line indicates the cosmological constant
model, $\rho_X(z)/\rho_X(0)=1$.
(a) Reconstructed $\rho_X(z)$ using 192 (shaded regions), 193 (dotted lines), and 
194 (dashed lines) SNe Ia from the Tonry/Barris
sample \citep{Tonry03,Barris03}.
(b) Reconstructed $\rho_X(z)$ using 192 Tonry/Barris sample SNe Ia   
(shaded regions) and 58 Knop sample SNe Ia (dotted lines)
\citep{Knop03}.}


\begin{thebibliography}{}


\bibitem[Aguirre(1999)]{Aguirre99}
Aguirre, A.N. 1999, ApJ, 512, L19


\bibitem[Alam et al.(2003)]{Alam03}
Alam, U.; Sahni, V.; Saini, T.D.; \& Starobinsky, A.A. 2003,
astro-ph/0311364

\bibitem[Albrecht et al.(2002)]{Albrecht02}
Albrecht, A.; Burgess, C.P.; Ravndal, F.; \& Skordis, C. 2002,
Phys. Rev. D65, 123507

\bibitem[Armendariz-Picon, Mukhanov, \& Steinhardt(2000)]{Armenda00}
Armendariz-Picon, C., Mukhanov, V., Steinhardt, P.J. 2000,
Phys. Rev. Lett. 85, 4438

\bibitem[Barger \& Marfatia(2001)]{Barger}
Barger, V., \& Marfatia, D. 2001, Phys. Lett. B, 498, 67

\bibitem[Barris et al.(2003)]{Barris03}
Barris, B.J., et al. 2003, astro-ph/0310843, ApJ, in press

\bibitem[Bean \& Melchiorri(2002)]{Bean02}
Bean, R.; \& Melchiorri, A. 2002,
Phys. Rev. D65, 041302

\bibitem[Bennett et al.(2003)]{Bennett03}
Bennett, C., et al. 2003, ApJ, Suppl. 148, 1

\bibitem[Bernstein \& Jain(2003)]{Bern03}
Bernstein, G.M.; \& Jain, B. 2003,
astro-ph/0309332, ApJ, in press

\bibitem[Bond, Efstathiou, \& Tegmark(1997)]{Bond97}
Bond,J.R.; Efstathiou, G.; \& Tegmark, M. 1997, MNRAS, 291, L33

\bibitem[Boyle, Caldwell, \& Kamionkowski(2002)]{bo02}
Boyle,  L.A.; Caldwell, R.R.; Kamionkowski, M. 2002,
Phys. Lett. B545, 17
 
 
\bibitem[Branch(1998)]{Branch98}
Branch, D. 1998, ARA\&A, 36, 17


\bibitem[Caldwell, Dave, \& Steinhardt(1998)]{caldwell98} Caldwell, R., Dave, R.,
Steinhardt, P. 1998, Phys. Rev. Lett., 80,  1582 

\bibitem[Carroll, Hoffman, \& Trodden(2003)]{Carroll03}
Carroll, S.M.; Hoffman, M., \& Trodden, M. 2003,
Phys. Rev. D68, 023509

\bibitem[Choudhury \& Padmanabhan(2003)]{Choud03}
Choudhury, T.R., \& Padmanabhan, T. 2003,
astro-ph/0311622 

\bibitem[Daly \& Djorgovski(2003)]{Daly03}
Daly, R.A., \& Djorgovski, S.G. 2003, ApJ, 597, 9


\bibitem[Deffayet(2001)]{ddg} Deffayet, C. 2001, Phys. Lett. B502, 199

\bibitem[Dodelson, Kaplinghat, \& Stewart(2000)]{Dode00}
Dodelson, S., Kaplinghat, M., \& Stewart, E. 2000, 
Phys. Rev. Lett. 85,  5276

\bibitem[Drell, Loredo, \& Wasserman(2000)]{Drell00}
Drell, P.S.; Loredo, T.J.; \& Wasserman, I. 2000, 
\apj, 530,  593

\bibitem[Farrar \& Peebles(2003)]{farrar03}
Farrar, G.R., \& Peebles, P. J. E. 2003, astro-ph/0307316

\bibitem[Freedman et al.(2001)]{Freedman01}
Freedman, W. L. et al. 2001, ApJ, 553, 47

\bibitem[Freese et al.(1987)]{fafm} Freese, K., Adams, F.C., Frieman, J.A.,
 and Mottola, E. 1987, Nucl. Phys. B287, 797

\bibitem[Freese \& Lewis(2002)]{freeselewis} Freese, K., and Lewis, M.,
2002, Phys. Lett., B540, 1


\bibitem[Frieman et al.(1995)]{frieman} Frieman, J.,  Hill, J., Stebbins, A.,
and Waga, I. 1995, Phys. Rev. Lett., 75, 2077


\bibitem[Frieman(1997)]{Frieman97} 
Frieman, J. A. 1997, Comments Astrophys., 18, 323

\bibitem[Gerke \& Efstathiou(2002)]{Gerke02}
Gerke, B.F., Efstathiou, G. 2002, MNRAS, 335, 33

\bibitem[Griest(2002)]{Griest02}
Griest, K. 2002, Phys. Rev. D66, 123501

\bibitem[Hamuy et al.(1996)]{H96}
Hamuy, M., et al. 1996, AJ, 112, 2408

\bibitem[Hannestad \& Mortsell(2002)]{Hanne02}
Hannestad, S., \& Mortsell, E. 2002, 
Phys. Rev. D66, 063508

\bibitem[Hawkins et al.(2003)]{Hawkins02}
Hawkins, E. et al. 2003, astro-ph/0212375, MNRAS in press

\bibitem[Holz \& Wald(1998)]{HolzWald98}
Holz, D.E. \& Wald, R.M. 1998, Phys. Rev., D58, 063501

\bibitem[Hu(2002)]{Hu02}
Hu, W. 2002, Phys.Rev. D66, 08351

\bibitem[Huterer \& Ma(2003)]{Huterer03}
Huterer, D., Ma , C. 2003, astro-ph/0307301 

\bibitem[Huterer \& Turner(2001)]{Huterer01}
Huterer, D., \& Turner, M.S. 
Phys. Rev. D64, 123527
 
 
\bibitem[Jimenez(2003)]{Jimenez03}
Jimenez, R. 2003, New Astron. Rev. 47, 761


\bibitem[Kantowski, Vaughan, \& Branch(1995)]{Kantowski95}
Kantowski, R.; Vaughan, T.; Branch, D. 1995, ApJ, 447, 35

\bibitem[Knop et al.(2003)]{Knop03}
Knop, R. A., et al. 2003, astro-ph/0309368, ApJ, in press

\bibitem[Kujat et al.(2002)]{Kujat02}
Kujat, J.; Linn, A.M.; Scherrer, R.J.; \& Weinberg, D.H. 2002,
ApJ, 572, 1

\bibitem[Kuo et al.(2002)]{ACBAR}
Kuo, C.L., et al. 2002, submitted to ApJ, astro-ph/0212289

\bibitem[Lewis \& Bridle(2002)]{LB02}
Lewis, A., \& Bridle, S. 2002, Phys. Rev. D, 66, 103511, astro-ph/0205436

\bibitem[Maor, Brustein, \& Steinhardt(2001)]{MBS00}
Maor, I., Brustein, R., \& Steinhardt, P.J. 2001, 
Phys. Rev. Lett., 86, 6; Erratum-ibid. 87 (2001) 049901

\bibitem[Maor et al.(2002)]{Maor02}
Maor,I.; Brustein, R.; McMahon, J.; \& Steinhardt, P.J. 2002, 
Phys. Rev. D65, 123003

\bibitem[Melchiorri et al.(2002)]{Mel02}
Melchiorri, A.; Mersini, L.; \"{O}dman, C.J.; \& Trodden, M. 2002,
astro-ph/0211522

\bibitem[Metcalf \& Silk(1999))]{ms99}
Metcalf, R. B., \&
Silk, J. 1999, ApJ, 519, L1

\bibitem[Majumdar \& Mohr(2003)]{Maju03}
Majumdar, S., \& Mohr, J.J. 2003,
astro-ph/0305341, submitted to ApJ

\bibitem[Mukherjee et al.(2003)]{Mukherjee03}
Mukherjee, P.; Banday, A.J.; Riazuelo, A.; Gorski, K.M.; \& Ratra, B. 2003,
astro-ph/0306147, ApJ, in press

\bibitem[Munshi, Porciani, \& Wang(2003)]{Munshi03b} 
Munshi, D.; Porciani, C.; \& Wang, Y. 2003,
astro-ph/0302510, MNRAS, in press


\bibitem[Munshi \& Wang(2003)]{Munshi03} 
Munshi, D., and Wang, Y. 2003, ApJ, 583, 566 

\bibitem[Neil(1993)]{neil}
Neil, R.M. 1993, ftp://ftp.cs.utoronto.ca/pub/~radford/review.ps.gz

\bibitem[\"{O}dman at el.(2002)]{Odman02}
\"{O}dman, C.J.; Melchiorri, A.; Hobson, M.P.; \& Lasenby, A.N. 2002,
astro-ph/0207286

\bibitem[Padmanabhan(2003)]{Pad03}
Padmanabhan, T. 2003, Physics Reports 380, 235-320 

\bibitem[Peebles \& Ratra(1988)]{peebles88}
Peebles, P. J. E.; Ratra, B. 1988, ApJ, 325L, 17

\bibitem[Peebles \& Ratra(2003)]{peebles03}
Peebles, P. J. E.; Ratra, B. 2003, Rev.Mod.Phys. 75, 559-606

\bibitem[Pearson et al.(2003)]{CBI} 
Pearson, T.J., et al. 2003, ApJ, 591, 556-574

\bibitem[Percival et al.(2002)]{Percival02}
Percival, W.J. et al. 2002, MNRAS, 337, 1068

\bibitem[Perlmutter et al.(1999)]{Perl99} Perlmutter, S., et al. 1999,  
ApJ,  517,  565


\bibitem[Phillips(1993)]{Phillips93}
Phillips, M.M., ApJ, 413, L105 (1993)

\bibitem[Podariu \& Ratra(2001)]{Poda01}
Podariu,  S.; \& Ratra, B. 2001,
ApJ, 563, 28


\bibitem[Press et al.(1994)]{Press94}
Press, W.H., Teukolsky, S.A., Vettering, W.T., 
\& Flannery, B.P. 1994, Numerical Recipes, Cambridge University Press, Cambridge.

\bibitem[Riess et al.(1998)]{Riess98} Riess, A.~G., et al. 1998,  
AJ, 116, 1009 

\bibitem[Riess, Press, \& Kirshner(1995)]{Riess95}
Riess, A.G., Press, W.H., and Kirshner, R.P., ApJ, 438, L17 (1995)

\bibitem[Riess et al.(1999)]{Riess99}
Riess, A.G., et al. 1999, AJ, 118,  2675

\bibitem[Sahni \& Shtanov(2002)]{Sahni02}
Sahni, V., \& Shtanov, Y. 2002,
astro-ph/0202346 
  

\bibitem[Schulz \& White(2001)]{Schulz01}
Schulz, A.E.; \& White, M. 2001,
Phys. Rev. D64, 043514


\bibitem[Seo \& Eisenstein(2003)]{Seo03}
Seo, H.; \& Eisenstein, D.J. 2003, astro-ph/0307460

\bibitem[Sereno(2002)]{Sereno02}
Sereno, M. 2002, Astron. Astrophys., 393, 757

\bibitem[Spergel et al.(2003)]{Spergel03}
Spergel, D.N., et al. 2003, astro-ph/0302209

\bibitem[Tegmark(2002)]{Tegmark02}
Tegmark, M. 2002, Phys. Rev. D66, 103507


\bibitem[Tonry et al.(2003)]{Tonry03}
Tonry, J.L., et al. 2003, ApJ, 594, 1-24


\bibitem[Verde et al.(2002)]{Verde01}
Verde, L. et al. 2002, MNRAS, 335, 432

\bibitem[Viel et al.(2003)]{Viel03}
Viel, M.; Matarrese, S.; Theuns, T.; Munshi, D.; \& Wang, Y. 2003,
MNRAS, 340, L47

\bibitem[Wambsganss et al.(1997)]{Wamb97}
Wambsganss, J., Cen, R., Xu, G., \& Ostriker, J.P. 1997,
ApJ, 475, L81

\bibitem[Wang(1999)]{Wang99} 
Wang, Y. 1999, ApJ, 525, 651 

\bibitem[Wang(2000a)]{Wang00a} 
Wang, Y. 2000a, ApJ, 531, 676 

\bibitem[Wang(2000b)]{Wang00b}
Wang, Y. 2000b, ApJ, 536, 531 

\bibitem[Wang \& Garnavich(2001)]{Wang01a}
Wang, Y., and Garnavich, P. 2001, ApJ, 552, 445 

\bibitem[Wang \& Lovelace(2001)]{Wang01b} 
Wang, Y., and Lovelace, G. 2001, ApJ, 562, L115 

\bibitem[Wang, Holz, \& Munshi(2002)]{Wang02} 
Wang, Y., Holz, D.E., and Munshi, D. 2002, ApJ, 572, L15 

\bibitem[Wang et al.(2003)]{Wang03}
Wang, Y.;  Freese, K.; Gondolo, P.; \& Lewis, M. 2003,
ApJ, 594, 25

\bibitem[Wang et al.(2004)]{Wang04}
Wang, Y., et al. 2004, in preparation

\bibitem[Wasserman(2002)]{Wass02}
Wasserman, I. 2002, Phys. Rev. D66, 123511

\bibitem[Weller \& Lewis(2003)]{Weller03}
Weller, J., \&  Lewis, A.M. 2003, MNRAS, in press


\bibitem[Zhu \& Fujimoto(2003)]{Zhu03}
Zhu, Z.; \& Fujimoto, M. 2003, astro-ph/0312022, ApJ in press

\end{thebibliography}
\end{document}